%% file: paper.tex
\documentclass[sigconf,screen]{acmart}

\usepackage[english]{babel}
\usepackage{graphicx}
\usepackage{grffile}
\usepackage{xspace}
\usepackage[normalem]{ulem}
\usepackage{listings}
\usepackage{enumitem}
\usepackage{colortbl}

\newcommand{\pathfigures}{figures/}

\newcommand{\ie}{i.e.\xspace}
\newcommand{\eg}{e.g.\xspace}
\newcommand{\etal}{\emph{et~al.}\xspace}

\newcommand{\databeacon}{$d_{beacon}$\xspace}
\newcommand{\datahist}{$d_{hist}$\xspace}

\newcommand{\datanow}{$d_{mar20}$\xspace}

\newcommand{\ripe}{RIPE\xspace}
\newcommand{\rviews}{RouteViews\xspace}

\definecolor{Gray}{gray}{0.8}

\copyrightyear{2020}
\acmYear{2020}
\setcopyright{none}
\acmConference[CoNEXT '20]{The 16th International Conference on emerging Networking EXperiments and Technologies}{December 1--4, 2020}{Barcelona, Spain}
\acmBooktitle{The 16th International Conference on emerging Networking EXperiments and Technologies (CoNEXT '20), December 1--4, 2020, Barcelona, Spain}
\acmPrice{15.00}
\acmDOI{10.1145/3386367.3432731}
\acmISBN{978-1-4503-7948-9/20/12}

\newcommand{\ext}[1]{}

\begin{document}

\title{Keep your Communities Clean: Exploring the Routing Message Impact of BGP Communities}

\author{Thomas Krenc}
\affiliation{
	\institution{Naval Postgraduate School}
}
\email{tkrenc@nps.edu}

\author{Robert Beverly}
\affiliation{
	\institution{Naval Postgraduate School}
}
\email{rbeverly@nps.edu}

\author{Georgios Smaragdakis}
\affiliation{
	\institution{TU Berlin}
}
\email{georgios.smaragdakis@tu-berlin.de}

\input{\pathsections abstract}



\maketitle

\vspace{-0.0em}
\input{\pathsections introduction}
\input{\pathsections background}
\input{\pathsections lab}

\input{\pathsections datasets}
\input{\pathsections methodology}

\input{\pathsections analysis}

\input{\pathsections discussion}

\section*{Acknowledgments}
\vspace{-1mm}
We want to express our gratitude for the support and feedback received from the
networking community via mailing lists and private communication, in particular
Randy Bush,
Greg Hankins,
Jakob Heitz,
Maria Matějka,
Donald Sharp,
Henk Smit
and Stefan Wahl.
We thank the reviewers and anonymous shepherd for their constructive
input and guideance.  This work supported in part by NSF grant
CNS-1855614, the European Research Council (ERC)
Starting Grant ResolutioNet (ERC-StG-679158),
by the German Ministry for Education and Research (BMBF) as BIFOLD - Berlin Institute for the Foundations of Learning
and Data (01IS18025A, 01IS18037A), 
and performed while the first author held an NRC Research Associateship award at the Naval Postgraduate School.
Views and conclusions are those of the authors and
should not be interpreted as representing the official policies or
position of the U.S.\ government, the NSF, ERC, or BMBF.

\balance
\bibliographystyle{ACM-Reference-Format}
\bibliography{paper}

\end{document}

%% file: sections/abstract.tex
\begin{abstract}

BGP communities are widely used to tag prefix aggregates for
policy, traffic engineering, and inter-AS
signaling.  Because individual ASes define their own community
semantics, many ASes blindly propagate communities they do not
recognize.  Prior research has
shown the potential security vulnerabilities when
communities are not filtered.
This work sheds light on
a second unintended side-effect of communities and permissive
propagation: an increase in \emph{unnecessary} BGP routing
messages.
Due to its transitive property, a change in the community attribute
induces update messages throughout established routes, just updating
communities.
We ground our work by characterizing the handling of
updates with communities, including when filtered, on multiple
real-world BGP implementations in
controlled laboratory experiments.
We then examine 10 years of BGP messages observed in the wild at two route collector systems.
In 2020, approximately 25\% of all announcements modify the community attribute, but
retain the AS path of the most recent announcement; an additional 25\%
update neither community nor AS path.
Using predictable beacon prefixes,
we demonstrate that communities lead to an increase in update messages
both
at the tagging AS and at neighboring ASes that neither add nor filter
communities.  This effect is prominent for geolocation communities
during path exploration:
on a single day, 63\% of all unique community attributes are revealed
exclusively due to global withdrawals.
\end{abstract}

%% file: sections/introduction.tex
\section{introduction}

Border Gateway Protocol (BGP), the Internet's inter-domain routing protocol, is
fundamental to the operation, policies, and economics of the network.
Unsurprisingly, the real-world behavior of BGP has been subject to intense 
scrutiny~\cite{griffin1999analysis,10.1145/248156.248160,roughan201110}. As an
extensible protocol, BGP and its usage have evolved in response to operator needs.
BGP communities \cite{rfc1997,BenoitOnBGPCommunities} are one such
example of an optional attribute that adds new meta-information
to routing messages.

Communities
are used to conveniently tag (an aggregate of) prefixes to enable particular
actions or policies.  However, community values and semantics are not
standardized -- the meaning of a specific community value is defined by
individual autonomous systems (ASes).  Benoit \etal first defined a taxonomy where, broadly speaking,
communities are used to tag received prefixes to aid an AS's internal routing
decisions, or added to outbound announcements as a convenient signaling
mechanism~\cite{BenoitOnBGPCommunities}.  Today, communities are known to encode
location
identifiers~\cite{Inferring-Complex-AS-Relationships:IMC2014,CoNEXT2015-P2F}
express policy preferences
upstream~\cite{IXP-communities-BGP-passive:Globecom2013},
enable selective advertisement in Internet Exchange Points
(IXPs)~\cite{IMC2014-RS,Inferring-Multilateral-Peering:CONEXT2013}, and as a
DDoS
mitigation signal~\cite{IMC2017-blackholing,Blackholing:PAM2016} (\eg BGP
blackholing~\cite{rfc7999,Cisco-remotely-triggered-black-hole-filtering}).

BGP communities have been widely adopted.  Streibelt \etal found a
250\% increase in the number of unique communities and a 200\% increase in the
number of ASes that
use communities as seen in BGP advertisements between 2010 and 2018~\cite{IMC2018-Communities-attacks}.
Giotsas \etal~\cite{SIGCOMM2017-peering-infrastrutcure-outages} report that around 50\% of IPv4 and 30\% of
IPv6 BGP announcements in 2016 include at least one location-tagged BGP community.

Despite the rich literature on BGP and BGP communities,
prior work has not investigated the unintended impact of communities on the
\emph{volume} of BGP message traffic in the wild.
Isolating and decoupling the traffic impact of communities from other
mechanisms and variables in the complex Internet without ground-truth
configurations is challenging.
In this work, we take a first step towards this larger goal by examining
changes in AS paths and communities in billions of update messages at 500+ peers over 10 years.
We find that updates with no path change are common throughout the entire
measurement period and are rooted in widespread community deployment,
increasingly interconnected networks, and lack of community filtering.
More specifically, we find:

\begin{enumerate}

\item Around 50\% of announcements in March 2020 signal no path change, while
half of these exhibit a community change. We consider these announcements
\textit{unnecessary}. %

\item In laboratory experiments and in the wild, we show that community
geo-tagging in combination with missing or ineffective filtering can lead to an
increase of update messages.

\item Among 10 tested router implementations, all but two forward
communities by default. Also, 7 routers generate duplicates due to filtering
communities at egress.

\item By utilizing beacon prefixes we show that more than 60\% of all encoded
information in community attributes is revealed during global withdrawals,
as a result of path exploration.

\end{enumerate}

We publicize and validate our work through
mailing lists and online documentation~\cite{url:communityexploration}. Our
findings resonated with the community, in particular with router
developers, and helped identify a source of duplicates in a current
popular BGP implementation.
For reproducibility we publish code to identify unnecessary
announcements and router configuration of all tested routers.
Our findings afford a better understanding of routing instabilities in the
Internet, unnecessary load on the system, and may help foster detection of
anomalous communities in the future. We conclude by discussing 
implications on routing
message archival and suggest future work.

%% file: sections/background.tex
\section{Background and Related Work}
\label{sec:background}

This section provides details
of BGP relevant to understanding our work on communities, as well as
a summary of
prior research.  We assume working familiarity with BGP;
see~\cite{roughan201110} for a broad overview.

\vspace{1mm}
\noindent \textbf{Path Exploration:}
The BGP decision process is
complicated, involving iBGP, eBGP and IGP interaction,
is governed by individual network policies, and beholden to BGP
implementation particulars.  There is a basic tension between
propagating reachability information quickly and sending
it prematurely, \ie, before the AS has converged on a new best
state.  %
In practice, implementations and configurations of BGP often delay
sending messages.
Thus, updates often occur in bursts.

Several prior works analyze network stability, path exploration, and
the trade off in withholding updates~\cite{10.1145/248156.248160}.  Mechanisms such as route
dampening and MRAI timers~\cite{deshpande2004impact} have been
explored, but may offer sub-optimal performance in reacting to routing
events~\cite{wang2006measurement}.  Thus, these mechanisms are
selectively deployed.
Indeed, we show that path exploration, combined
with BGP community use, is a significant contributor to BGP update
traffic.

\vspace{1mm}
\noindent \textbf{Communities:}
BGP messages are relatively simple and include prefix \emph{updates}
(often termed an ``announcement'') as well as prefix
\emph{withdrawals} (indicating that the prefix should be removed from
the routing table).  In addition, BGP messages can include multiple
optional attributes, among them the next-hop, MED, and community.
Among these, BGP communities are notable because they are transitive
-- meaning that they are an optional attribute that may be
propagated.  Communities are the focus of our study.  As we will
show, not only are BGP communities in common use, but they are a
primary contributor to overall BGP message traffic.

BGP communities are simply a 32-bit value, however a common convention
is that the upper two bytes encode the ASN of the AS that ``owns'' the
community, while the lower two bytes define the meaning of the
community.  Because communities have no well-defined semantics, it is
up to each individual AS to define the meaning of the lower two bytes
corresponding to their community space.   Note that while large and
extended communities have since been added~\cite{rfc8092,rfc8195}, for
instance to accommodate 32-bit ASNs and to communicate additional bits
of information, these are in infrequent use.  Hence, our study focuses
on traditional BGP communities.

Benoit \etal provides a taxonomy of BGP communities
in~\cite{BenoitOnBGPCommunities}.  As a contemporary convention, BGP
communities can be broadly divided into informational communities and
action communities~\cite{rfc8195}.  Informational communities are
typically added to ingress routing announcements to tag aggregates of
routes in a common way in order for an AS to make internal policy and
routing decisions.  For instance, a common informational community
used by large ASes is to encode the physical geographic location
where a prefix is received,
\eg ``North
America, Dallas, TX.''

In contrast, action communities are frequently added to egress 
announcements to implement in-band signaling
to a different AS.
For instance, a common use of action communities is the blackhole
community which indicates that a provider should stanch traffic for a
particular IP or prefix that is experiencing a DDoS attack.

With the growth in BGP community adoption, researchers have
in recent years explored community prevalence and
security~\cite{BenoitOnBGPCommunities, IMC2018-Communities-attacks},
as well as the information they leak about connectivity, attacks, and
outages~\cite{SIGCOMM2017-peering-infrastrutcure-outages,
IMC2017-blackholing}.  However, less attention has been given to the
unintended impacts of communities on the volume of BGP message traffic
in general, and updates in particular -- these are the focus of the
present research.

\vspace{1mm}
\noindent \textbf{Duplicate Updates:} As with the protocol itself, BGP update behavior has been extensively studied, for
instance to understand routing dynamics~\cite{LiBGPRoutingDynamics},
understand convergence and forwarding behavior~\cite{MaoBGPBeacons},
quantify path performance~\cite{wang2006measurement}, and locate
origins of instabilities~\cite{FeldmannLocatingInstabilities}.  Of
most relevance to our present study are so-called ``duplicate'' BGP
updates, superfluous messages that do not update routing state. First
identified by Labovitz in 1998~\cite{labovitz1998internet}
and
believed to be attributable to buggy implementations, Park \etal
subsequently demonstrated that iBGP/eBGP interaction was the primary
cause of these duplicates~\cite{park2010investigating}.
As we also find,
when a router receives an internal update with a changed
attribute, that attribute may be removed or replaced prior to announcing to an
eBGP peer, resulting in a duplicate.
Hauweele \etal later verified in both real and
lab experiments that MED, next-hop, and community attribute changes induce these
duplicates~\cite{hauweele2018parrots}.

While duplicate updates have been a recognized issue for decades, we first show
via controlled lab experiments in~\S\ref{sec:controlledexperiments} the
propagation, update, and duplication implications of communities specifically.
Second, our work attempts to quantify the impact of communities on BGP message
traffic in the wild and over time. Our findings in~\S\ref{sec:analysis}
highlight the effects of increased BGP community use on update generation.
We show that BGP geolocation communities are a primary source of unnecessary
updates, and induce inter-AS message traffic even when communities are filtered.

%% file: sections/lab.tex
\section{Controlled Experiments}\label{sec:controlledexperiments}

To validate our findings and inferences, as well as gain a deeper
understanding of BGP update root causes,
we conduct a series of experiments in a
controlled laboratory setting.

\begin{figure}[t]
\center
\includegraphics[width=0.8\linewidth]{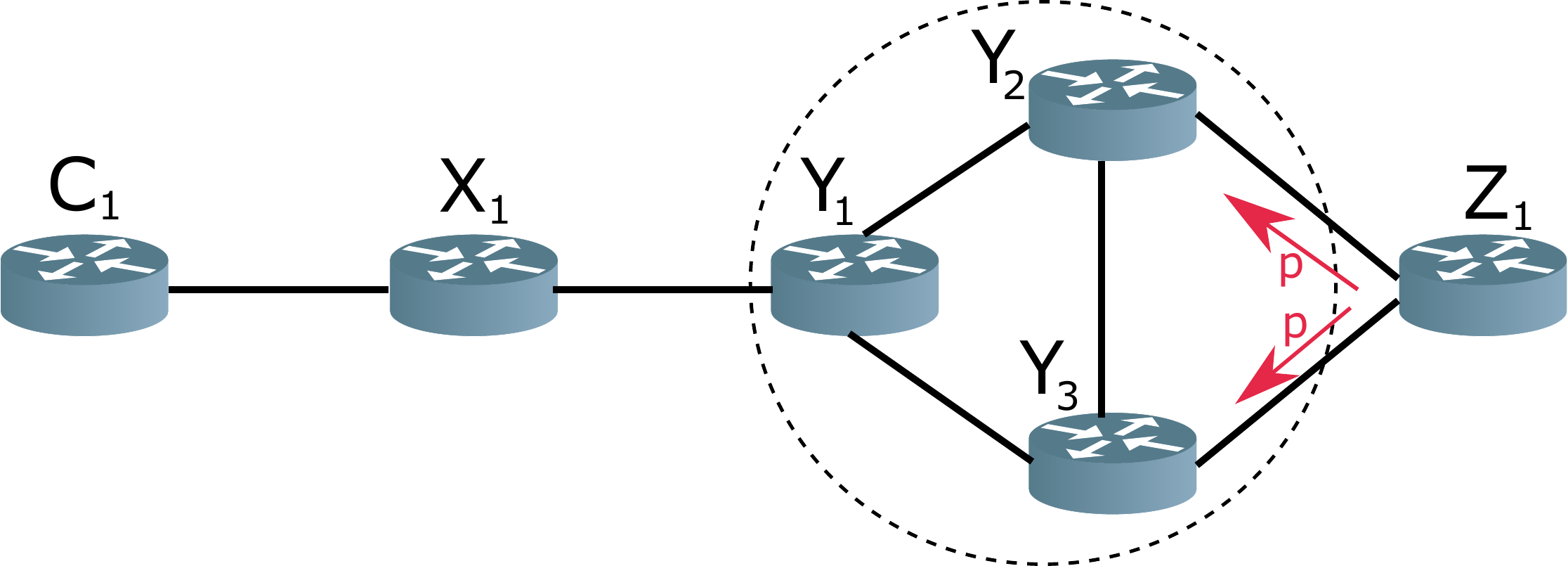}
\vspace{-3mm}
\caption{Laboratory topology to understand conditions generating
BGP update messages}
\label{fig:lab1}
\vspace{-4mm}
\end{figure}

For each experiment run, we configure all
routers depicted in Figure~\ref{fig:lab1} to use one of the following
routing software: Cisco IOS (12.4(20)T), IOS XR (v6.0.1),
Juniper Junos (Olive 12.1R1.9),
Nokia SR OS (20.7.R2),
BIRD (v1.6.6 and v2.0.7)~\cite{url:bird},
FRRouting (v6.0.2)~\cite{url:frrouting},
OpenBGPD (v5.2 and v6.6)~\cite{url:openbgpd}
or Quagga (v1.2.4)~\cite{url:quagga}.
While Cisco and Juniper routers dominate the core router market, for example
BIRD is used in many large IXPs \cite{IMC2014-RS,euroix-2015}.
FRRouting and Quagga are used in \rviews and \ripe collectors to
listen for BGP updates and dump them to Multi-Threaded Routing Toolkit
(MRT)~\cite{rfc6396} files. Our list of routing software is not comprehensive.

We run IOS, Junos and SR OS in emulation. By using these router images,
as well as routing daemons, we can gain insight into real-world BGP
implementation behavior.
Because we find identical behaviors for most of the experiments, we report only
on the common behavior and where it deviates.

The lab topology is crafted to test several scenarios, with and
without communities and filtering, to understand the conditions that generate BGP
update messages and when those update messages are propagated.  The
topology consists of four ASes: $X,Y,Z$ and $C$.  Router
$C_1$ mimics a route collector, while $Z_1$ originates the prefix $p$.
The links in the topology correspond to both physical connections and
eBGP and iBGP sessions.  AS $Y$ has three routers within its network;
both $Y_2$ and $Y_3$ peer with AS $Z$.

Prior to running our experiments, we verify that only BGP keep alive
messages, \ie, pairwise heartbeats to test liveness, are sent once the network
has converged.
In the following experiments, we are interested in BGP messages that carry
announcements and withdrawals.
We use tools like tcpdump and generated logs to inspect this message exchange.

\begin{itemize}[leftmargin=*]
\item \textbf{Exp1}: We begin without any BGP communities to characterize
default behavior.
Note that border router $Y_1$ has two paths to reach $p$.  In the
absence of any policy, the BGP tie breaker selects $Y_2$ as the next
hop.  Therefore, to induce BGP updates, we disable the $Y_1$ to $Y_2$
link, and perform a packet capture of all messages arriving at the
collector $C_1$ and between $X_1$ and $Y_1$.

Without BGP communities, when $Y_1$ chooses a new next hop of $Y_3$,
it sends an update message to $X_1$ even though the AS path has not
changed.
(Note: Junos and SR OS do not generate duplicates).
However, this update is ignored by $X_1$ and
not propagated further -- no update message is observed at the
collector.

\item \textbf{Exp2}: Next, we consider the common scenario where AS $Y$ implements
communities that geographically tag incoming advertisements.  $Y_2$
adds community \texttt{Y:300} on ingress while $Y_3$ adds \texttt{Y:400}.
Because $Y_2$ is preferred, and no community filtering is implemented
in this network, the collector sees $p$ with \texttt{Y:300}.  We
again disable the $Y_1$ to $Y_2$ link.

Again, this induces an update
message from $Y_1$ to $X_1$. While the AS path is unchanged, this
update includes a changed community value of \texttt{Y:400}.  Because
the community value changed (implicit withdrawal), $X_1$ also sends an update which is
seen at the collector. Note that while updates sent by $Y_1$ can be due to an
internal next-hop change (as in Exp1), in the case of $X_1$ the next-hop does
not change. Thus, a change in the community attribute is the sole trigger
for the update. (Note: IOS needs explicit configuration to forward communities
on eBGP sessions).

\item \textbf{Exp3}: We implement community filtering on $X_1$ by configuring
it to remove all communities on egress.  We again flap the $Y_1$ to
$Y_2$ link to generate the update message.  Surprisingly, even
though $X_1$ is removing communities, it still sends an update to
the collector
(Note: Junos, SR OS and OpenBGPD v6.6 do not generate duplicates).
Note that this update has an unchanged AS path and
includes no communities -- \ie, it is an arguably unnecessary
message.

\item \textbf{Exp4}: We then repeat experiment 3, but modify $X_1$
to filter communities on ingress from $Y_1$.  In this case, the
spurious update message is not sent as the communities are not
contained in the router's RIB.  This shows that we can differentiate
between ingress and egress community filtering.
\end{itemize}

\footnotesize
\begin{table}
\caption{Overview of experiments and results.}
\vspace{-3mm}
\label{tab:labsummary}
\centering
\begin{tabular}{|c|p{1.2cm}|p{1.2cm}|p{1.2cm}|p{1.2cm}|}
\hline
Routing software
 & Exp1: $Y_1$ sends dups (next-hop change)
 & Exp2: $X_1$ forwards communities set by $Y_1$
 & Exp3: $X_1$ cleans at egress, sends dups
 & Exp4: $X_1$ cleans at ingress, no dups generated\\\hline
\hline
Cisco IOS & & & & \\
12.4(20)T & true & \textbf{false} & true & true \\
XR v6.0.1 & true & \textbf{false} & true & true \\\hline
Juniper Junos & & & & \\
Olive 12.1R1.9 & \textbf{false} & true & \textbf{false} & true \\\hline
Nokia SR OS & & & & \\
20.7.R2  & \textbf{false} & true & \textbf{false} & true \\\hline
BIRD & & & & \\
v1.6.6 & true & true & true & true \\
v2.0.7 & true & true & true & true \\\hline
FRRouting & & & & \\
v6.0.2 & true & true & true & true \\\hline
OpenBGPD & & & & \\
v5.2 & true & true & true & true \\
v6.6 & true & true & \textbf{false} & true \\\hline
Quagga & & & & \\
v1.2.4 & true & true & true & true \\
\hline
\end{tabular}
\end{table}
\normalsize

\noindent \textbf{Summary:} 
Table~\ref{tab:labsummary} summarizes the experiments and results.
Among the tested software, by default, only Junos and SR OS
prevent duplicates from being generated by, \eg internal changes or community
filtering on egress.
OpenBGPD v6.6 suppresses duplicates when the community changes but not when the
next-hop changes.
Furthermore, all routers generate updates that are triggered only due a change
in the community attribute, if communities are not filtered at ingress.
Our findings imply that this behavior is transitive.
For reproducibility, we publish the relevant configurations for
each tested router software~\cite{url:communityexploration}.

We note that sending updates with no changes contradicts BGP specifications.
According to RFC4271~\S9.2~\cite{rfc4271}: ``A BGP speaker SHOULD
NOT advertise a given feasible BGP route from its Adj-RIB-Out if it would
produce an UPDATE message containing the same BGP route as was previously
advertised.''
In reality, maintaining the Adj-RIB-Out requires keeping significant state
which 
is a major concern for operators when making tradeoffs
between convergence speed and resource usage.
Also, vendors and developers design their software with different default
configurations.
While for example Junos maintains the Adj-RIB-Out by default, BIRD requires
explicit configuration by the user.
The open-source implementations are not only feature rich, but also
they can operate on many different hardware architectures with 
different memory capacities.
Thus, default configurations are kept at a minimum setting leaving the
responsibility to the user.

%% file: sections/datasets.tex
\section{Data Sets}
\label{sec:datasets}

\begin{table}
\caption{Overview \datanow data set}
\label{tab:datasetsoverview}
\centering
\begin{tabular}{|l|r||l|r|}
\hline
IPv4 prefixes & 1,071,150 & Announcements & 1,008M  \\ 				%
IPv6 prefixes & 99,141 & w/ communities & 737.0M \\				%
ASes & 68,911 & ~~~uniq. 16 bits & 5,778 \\
Sessions & 1,504 & uniq. AS paths & 43.9M \\					%
Peers & 581 & Withdrawals & 38.5M \\						%
\hline
\end{tabular}
\end{table}

To study the impact of BGP communities on update message propagation
in the wild, we use publicly available archived routing traffic from
the \rviews~\cite{routeviews} and \ripe~RIS~\cite{riperis} BGP collector projects.
We obtain all MRT formatted update (251,493) and RIB (9,539) files from all
collectors, inclusive of both IPv4 and IPv6 prefixes as well as withdrawals,
for a full day every 3 months (2019-03-15, 2019-06-15, 2019-09-15, etc.)
across a ten-year span (2010 to 2020).
While our analyses are based on update messages only, we use the RIB snapshots
to detect peer ASes that do not occur in the update files.

Prior to analyzing the raw data, we first perform basic
filtering, cleaning, and normalization, so as to not impart unintentional
bias.
Using current and historical allocation information from the regional
registries, we remove messages that contain an unallocated ASN or prefix at
the time of the message.
We do not aggregate overlapping prefixes, and we keep all prefixes, regardless
of their length.
We note that 7 peer ASes do not prepend their ASN to the AS path when
advertising routes to the collector, i.e., the $FROM$ field in the update
message differs from the left-most AS in the AS path. Those ASes are IXP
route servers.
To avoid overcounting peer ASes and avoid ambiguity when processing the data,
we add the ASN of the route server to the AS path.
Finally, only 4 BGP collectors (\rviews: \texttt{route-views3, eqix,
linx, sfmix}) record update messages at a microsecond granularity, as of March
2015.
All other collectors use a single second granularity for received updates.
When multiple messages arrive within the same second for these collectors,
we preserve the message ordering and assume that each subsequent message arrives
1$\mu$s after the previous.

In the remainder of this paper, we refer to the resulting data set as \datahist.
We use \datanow to point to the most recent data in our measurements, which is March 15, 2020.
Table~\ref{tab:datasetsoverview} provides an overview of \datanow.
The data set includes 1,504 sessions across 581 unique peer ASes.
The number of BGP sessions at these two collector projects has roughly
doubled over the past ten years.
We note that not all the peers send updates on any day.
In \datanow, we find updates for 451 of the 581 peer ASes.

\textbf{Routing Beacons:}
Routing beacons are prefixes announced and
withdrawn at periodic intervals~\cite{MaoBGPBeacons}.
Beacons can help network operators and researchers investigate the
routing system and routing anomalies by providing
a source of predictable and known behavior.
\ripe operates routing beacons~\cite{url:risbeacons} with an
update pattern of a single announcement every 4 hours, starting at
00:00 UTC, and a single withdrawal every 4 hours, starting at 02:00 UTC.
One specific IPv4 and IPv6 beacon prefix is announced per \ripe route collector.

From 39 available \ripe beacon prefixes, we remove 4 IPv4 beacons that are
either not active or too noisy. Then, we select all announcements and withdrawals
that are associated with the remaining 35 beacons.
We observe 660,567 announcements and 115,892 withdrawals spread over 998
sessions, 354 peers, and 34 collectors.
We refer to this subset as \databeacon.

%% file: sections/methodology.tex
\section{Announcement Types}
\label{sec:updatecategories}

To better understand announcements in our studied data sets, we first group them
by the prefix and the BGP session (the \texttt{<peer AS, next-hop>} tuple), in arriving
order.
Then, from one announcement to the next, we look for changes (or no changes)
in the community attribute and in the AS path. In the AS path, we further
distinguish between a change in the set of ASNs, and a change in path prepending,
i.e., inflation or deflation of an identical set; it cannot be both.
From three possible observations in the AS path attribute and two in the
community attribute, we define six different combinations of two letters to
label the \textit{announcement type}:
The first letter indicates the AS path (\textit{p} = path change,
\textit{n} = no path change, \textit{x} = path prepending),
and the second letter indicates
the community attribute (\textit{c} = community change, \textit{n} =
no community change):
\textit{pc},
\textit{pn},
\textit{nc},
\textit{nn},
\textit{xc},
\textit{xn}.

An announcement with a path change only is in the category $pn$.
If there is a change in path prepending (the set of ASes are
equal), it is in $xn$.
If, in addition to the path also the community attribute changes, the
announcement is in $pc$ (or $xc$ in case of path prepending).
While we intuitively expect $pn, pc, xn$ and $xc$ updates,
we also see updates without a path change:
$nc$ and $nn$ cover all announcements with no path change,
while the former also includes changes in the community attribute.
We note that $nn$ also includes two empty community attributes in succession.
Also, we acknowledge that the MED attribute for a the \texttt{<peer AS, next-hop>}
tuple can change towards the collector
as a reason for an \textit{nn} announcement.

\textbf{Statistics:}
\begin{table}[t]
\caption{Announcement types (share in \datanow and \databeacon)}
\vspace{-3mm}
\label{tab:updateTypes.overview}
\centering
\begin{tabular}{|c|l|r|r|}
\hline
\textbf{type} & \textbf{observed changes} & \datanow & \databeacon \\\hline
\hline
$pc$ & path $+$ community & 33.7\% & 44.6\% \\
$pn$ & path only & 15.1\% & 29.9\% \\
\rowcolor{Gray}
\textbf{nc} & \textbf{community only} & \textbf{24.5\%} & \textbf{13.8\%} \\
\rowcolor{Gray}
\textbf{nn} & \textbf{no change} & \textbf{25.7\%} & \textbf{11.2\%} \\
$xc$ & path prepending $+$ comm. & 0.3\% & 0.2\% \\
$xn$ & path prepending only & 0.7\% & 0.3\% \\\hline
\end{tabular}
\vspace{-4mm}
\end{table}
Table~\ref{tab:updateTypes.overview} provides a break-down of the possible
observations in \datanow.
We note that $nc$ and $nn$ -- the only types that do not include a path change --
make up more than half of all announcements (24.5\% and 25.7\% respectively).
The largest group of announcements is of type \textit{pc} with a share of 33.7\%,
while \textit{pn} announcements constitute 15.1\%.
The number of announcements that either inflate or deflate a given AS path is
negligibly small, contributing around 1\%.
For comparison, we also provide the share of types in \databeacon. Here, we see
a different distribution. While \textit{nc} and \textit{nn} together contribute
25\% to all announcements, \textit{pc} is the most dominant type with a share
of 44.6\%, followed by \textit{pn} with 29.9\% share. Again, \textit{xn} and
\textit{xc} are low in numbers.
Recall that beacon prefixes provide us with a more controlled view on the
update behavior since they are announced and withdrawn at stated intervals.
In the wild, however, unpredictable changes at the origin AS can add to the
dynamics of update propagation at all downstream paths.

\begin{figure}[t]
\centering
\includegraphics[width=1.0\linewidth]{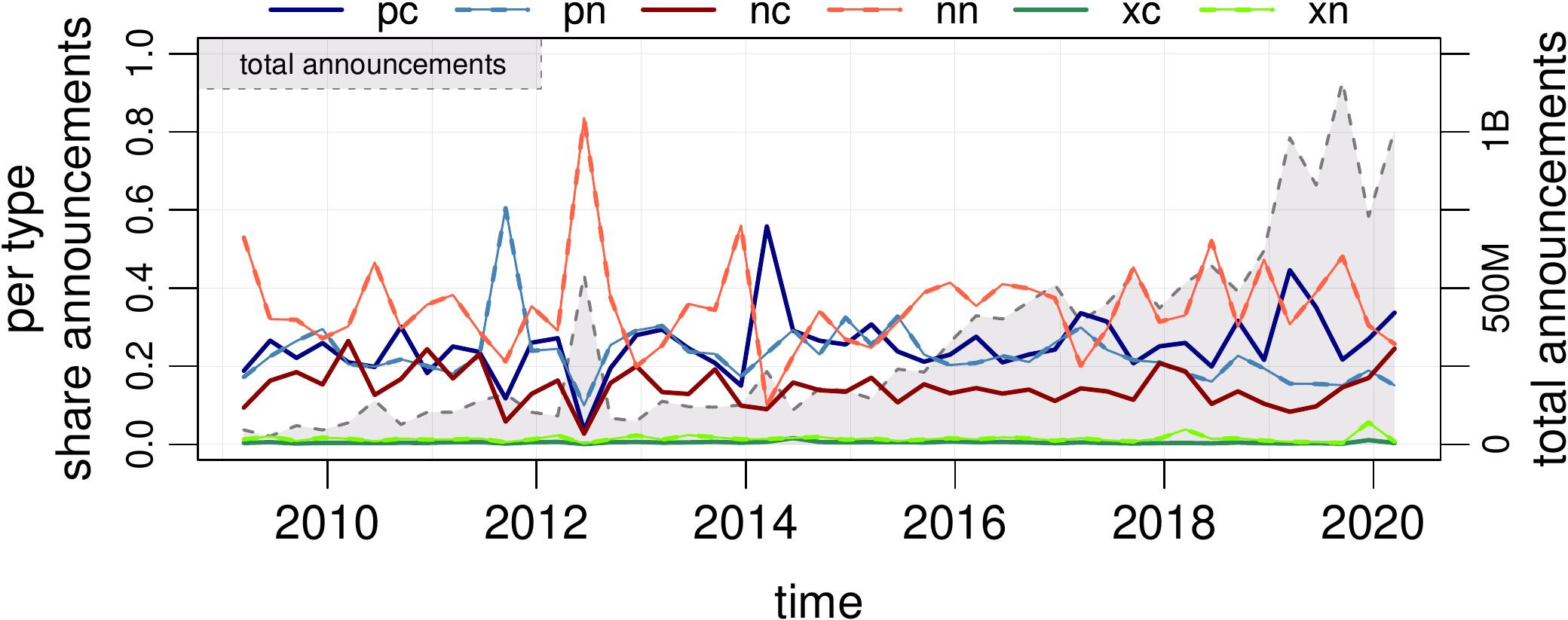}
\vspace{-1.8em}
\caption{Share of daily announcements per type and total announcements in \datahist.}
\label{fig:overTimeUpdateTypes.longall.share}
\end{figure}
Next, we investigate the longitudinal behavior of announcement types, using
\datahist.
In Figure~\ref{fig:overTimeUpdateTypes.longall.share}, we show the share
of the individual constituents on the left y-axis, and the total number of
announcements on the right y-axis, over time.
First, we note that the overall number of announcements show a 40-fold increase
from around 26M in 2009 to 1.1B in 2019.
Except for the spike\footnote{The spike of \textit{nn} activity in June 2012 is an
artifact of peer AS821 at the \texttt{route-views2} collector sending more than
1220 duplicate announcements for more than 210K prefixes throughout a single day.} in
2012, the number is below 250M until 2015, after which it roughly
doubles to 500M in 2018, followed by another doubling in 2019 to 1B. This
increase can be explained by the parallel increase of peer ASes that send
update messages to the collectors, but also by the overall increase of routing
activity.
Second, looking at the share of the individual announcement types, we observe
some amount of variability in the distribution over time.
The standard deviation over all time points ranges between 0.047 and 0.119 for
the individual types.
The median of (\textit{pc}, \textit{pn}, \textit{nc}, \textit{nn})
announcements over the last ten years is (26\%, 24\%, 15\%, 31\%), respectively.
We note that \datahist is limited to a full day of update data per quarter year.
In a separate analysis, using ten consecutive days in February 2020, we find
less variability ($\sigma < 0.04$) on a daily basis.

%% file: sections/analysis.tex
\section{Unnecessary Updates}
\label{sec:analysis}

Next, we study communities in
update messages with no path change and their impact on update propagation.
We focus on announcements and withdrawals for individual beacon prefixes
(\databeacon) visible in BGP sessions over 24 hours.

\begin{figure}[t]
\centering
\includegraphics[width=1.0\linewidth]{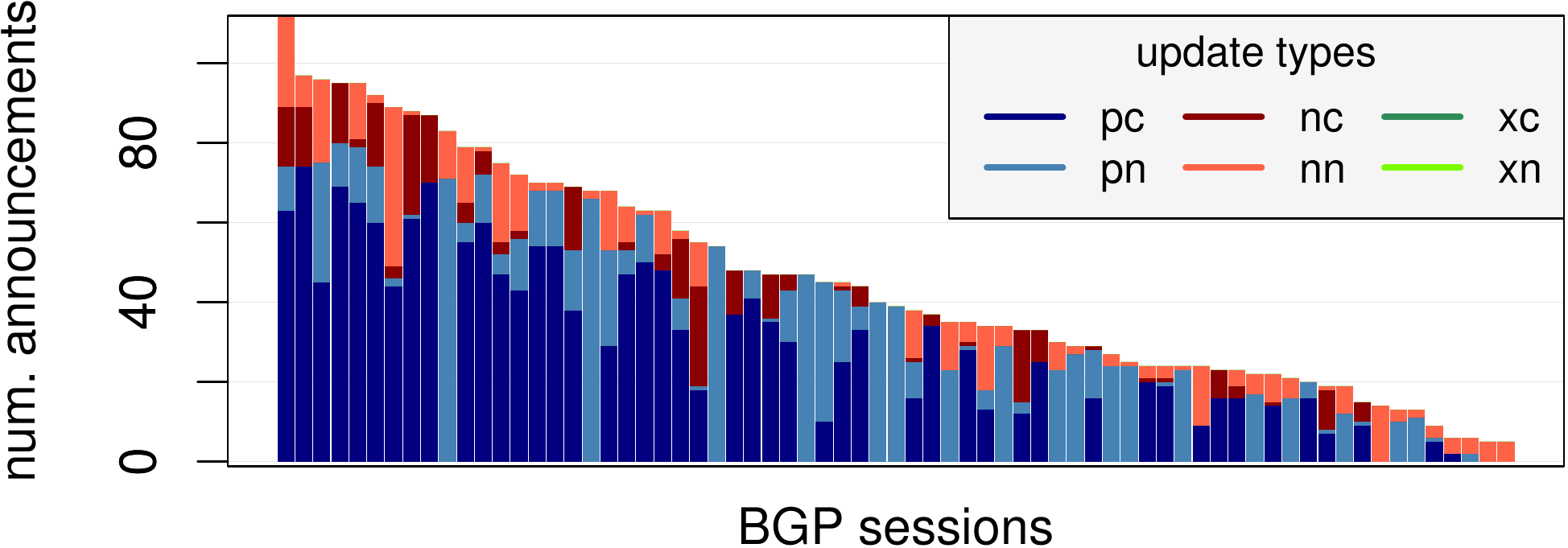}
\vspace{-5mm}
\caption{Announcement types per BGP session for beacon prefix $84.205.64.0/24$, collector: rrc00, March 15, 2020}
\label{fig:long-ext.20200315.rrc00.84-205-64-0-24.a1}
\end{figure}

\textbf{BGP Sessions.}
We begin by investigating how peers of a routing collector perceive
the different announcement types, \ie
\textit{pc}, \textit{pn}, \textit{nc}, \textit{nn}, \textit{xc}, and \textit{xn}.
We note that a peer AS sends only the best path via BGP sessions to the
collectors.
The stacked bar plot in Figure~\ref{fig:long-ext.20200315.rrc00.84-205-64-0-24.a1}
includes each of the BGP sessions of \ripe collector \texttt{rrc00}.
The sessions are sorted by number of announcements visible for prefix $84.205.64.0/24$~$\in$~\databeacon
and colored to indicate the announcement type.
We observe that each session shows a different
number of announcements. But more interesting is the fact that each
session shows a diverse distribution of announcements types, despite looking
\textit{only} at a single beacon prefix.
We characterize the peer ASes later in this section.
To this end, the root causes, i.e., why
\textit{nc} and \textit{nn} announcements are sent in the first place,
are
unclear. In the following we take a closer look at those announcement types.

\textbf{Community Exploration.}
To highlight the conditions that can lead to \textit{nc}
announcements in the wild,
we present an example using the view of a single BGP session.
\begin{figure}[t]
\centering
\includegraphics[width=1.0\linewidth]{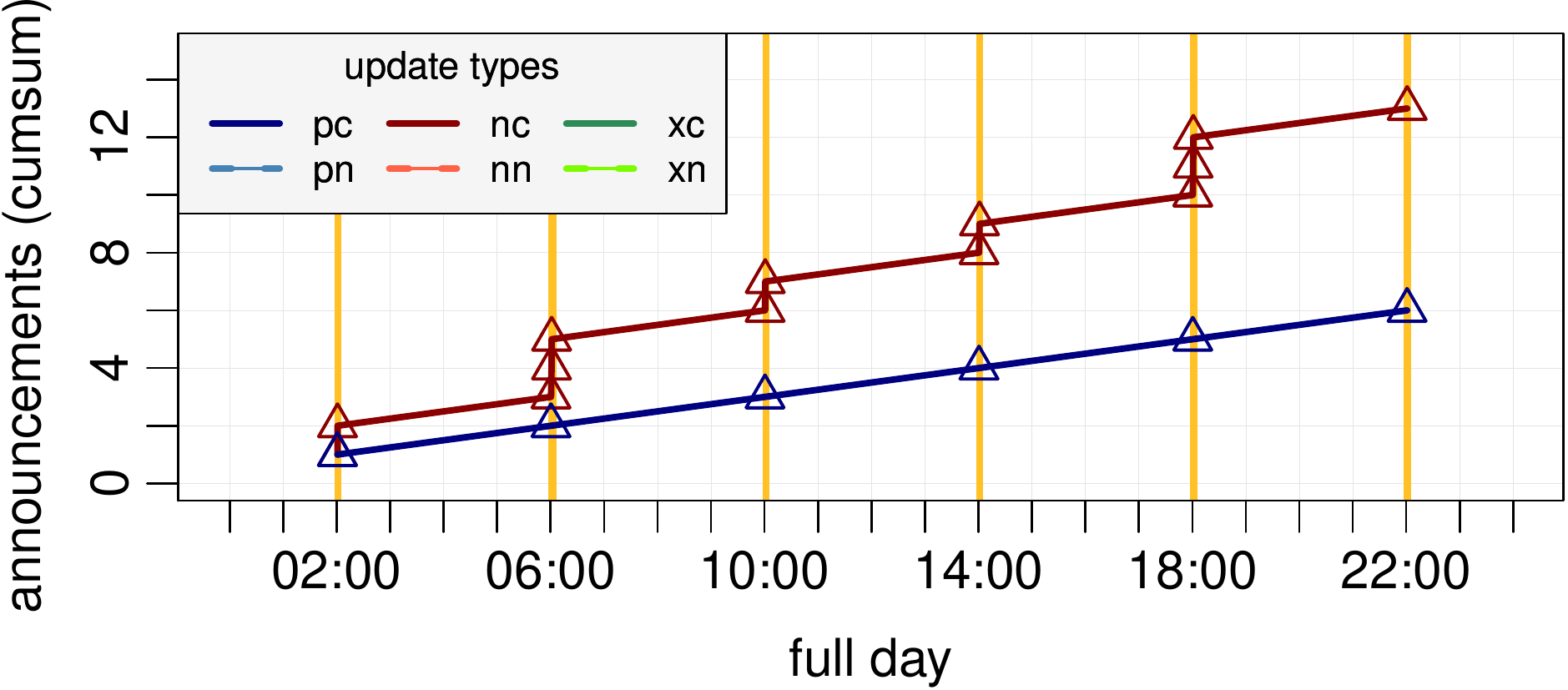}
\vspace{-5mm}
\caption{Announcement types over time with prefix $84.205.64.0/24$ and AS path (\texttt{20205 3356 174 12654}). Geo-tagging induces \textit{nc} announcements.}
\label{fig:long-ext.20200315.rrc00.84-205-64-0-24.c3.17.1}
\end{figure}
In Figure~\ref{fig:long-ext.20200315.rrc00.84-205-64-0-24.c3.17.1} we show the cumulative sum of announcements over 24 hours of March 15, 2020.
We plot all announcements for the same prefix $84.205.64.0/24$~$\in$~\databeacon via a single AS path (\texttt{20205 3356 174 12654}).
Vertical yellow lines indicate the arrival of a withdrawal message for that prefix, confirming the withdrawal interval for routing beacons.

All announcements for this particular route and day show up only
during the withdrawal phases, \ie at around 02:00, 06:00, 10:00, etc.
We deduce that this particular route was never a best path during that day (all time best path: \texttt{20205 6939 50304 12654}).
During the six withdrawal phases we observe a total of 19 announcements:
Starting with a \textit{pc} update (6 total), i.e., an announcement with changed path and community,
followed by multiple (13 total) \textit{nc}'s, announcements with changing community only.
Peer AS20205 does not set any communities.
However, it does not clean communities from its neighbor either:
The changing communities in \textit{nc} announcements represent
encoded ingress locations set presumably by AS3356.
We observe a total of 9 locations encoded in 19 announcements:
9 city communities, two country and two geographical regions, i.e., Europe and North America.
Per withdrawal phase, the location communities are mostly unique.

Due to distinct location communities attached to a single route,
multiple \textit{nc} announcements occur (comparable to Exp2 in
\S\ref{sec:controlledexperiments}). Analogously to path exploration,
we refer to this behavior as \textit{community exploration}:
Instead of multiple paths being announced, multiple communities for a single path are announced.
Also, the example above demonstrates that setting communities by one AS,
can impact the update behavior of a different AS, if no proper filtering is in
place (comparable to Exp4).

\textbf{Duplicate Announcements.}
Next, we explore a possible reason for the occurrence of \textit{nn} updates.
Therefore, we choose a route similar to the previous community exploration example.
However, we replace the peer AS with one that removes all communities (in $>$99\% of the cases).
\begin{figure}[t]
\centering
\includegraphics[width=1.0\linewidth]{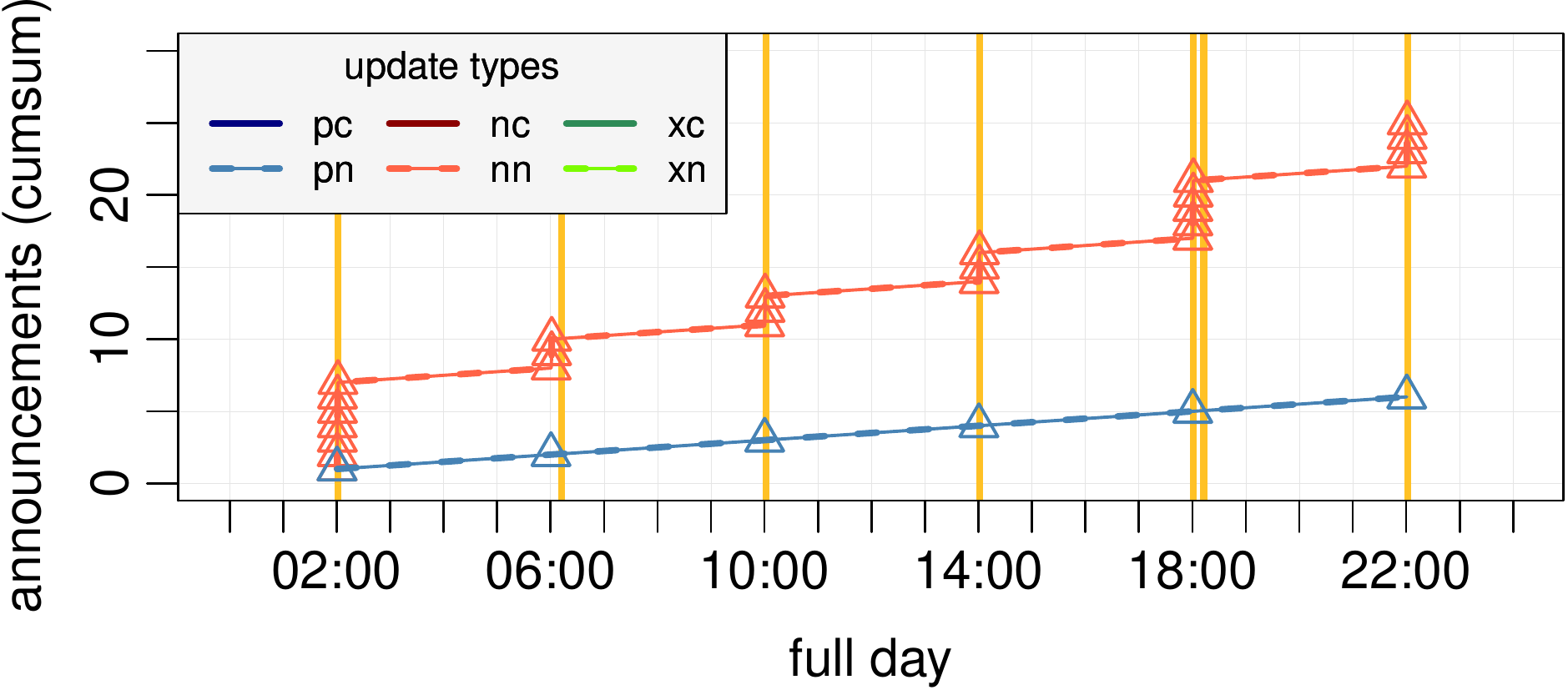}
\vspace{-3mm}
\caption{Announcement types over time with prefix $84.205.64.0/24$ and AS path (\texttt{20811 3356 174 12654}). Cleaning at egress generates \textit{nn} announcements.}
\label{fig:long-ext.20200315.rrc10.84-205-64-0-24.c3.2.1}
\vspace{-0mm}
\end{figure}
Figure~\ref{fig:long-ext.20200315.rrc10.84-205-64-0-24.c3.2.1} shows the cumulative sum of announcements over the day of March 15, 2020.
We plot announcements for the same prefix $84.205.64.0/24$, but via a different AS path (\texttt{20811 3356 174 12654}).
Vertical yellow lines represent withdrawal messages for that prefix, again in accordance with the predefined intervals.
Again, all 31 announcements occur during the withdrawal phases.
Also, the phases begin with a path change (6 total), here \textit{pn}, followed by
a series of \textit{nn} announcements (25 total).

Deduced from our previous observations, we speculate that during the withdrawal
phase AS20811 simply reannounces multiple \textit{nc}'s from AS3356 (as an
implicit withdrawal) and removes the existing communities prior to announcing,
thus inducing \textit{nn} announcements.
Note, we have demonstrated such behavior in lab experiments (Exp3).
We manually re-visit the raw BGP data and confirm that no other attribute towards the collector, e.g., the MED,
has changed and no other prefix is included in the updates.
However, since our observations are limited to inter-AS changes,
we do not exclude the possibility for other sources of \textit{nn}
announcements, \eg streams of updates due to intra-AS changes,
misconfiguration, or rate limiting -- in large and complex networks, the interplay
of these multiple internal factors may lead to unnecessary updates.

\begin{figure}[t]
\centering
\includegraphics[width=1.0\linewidth]{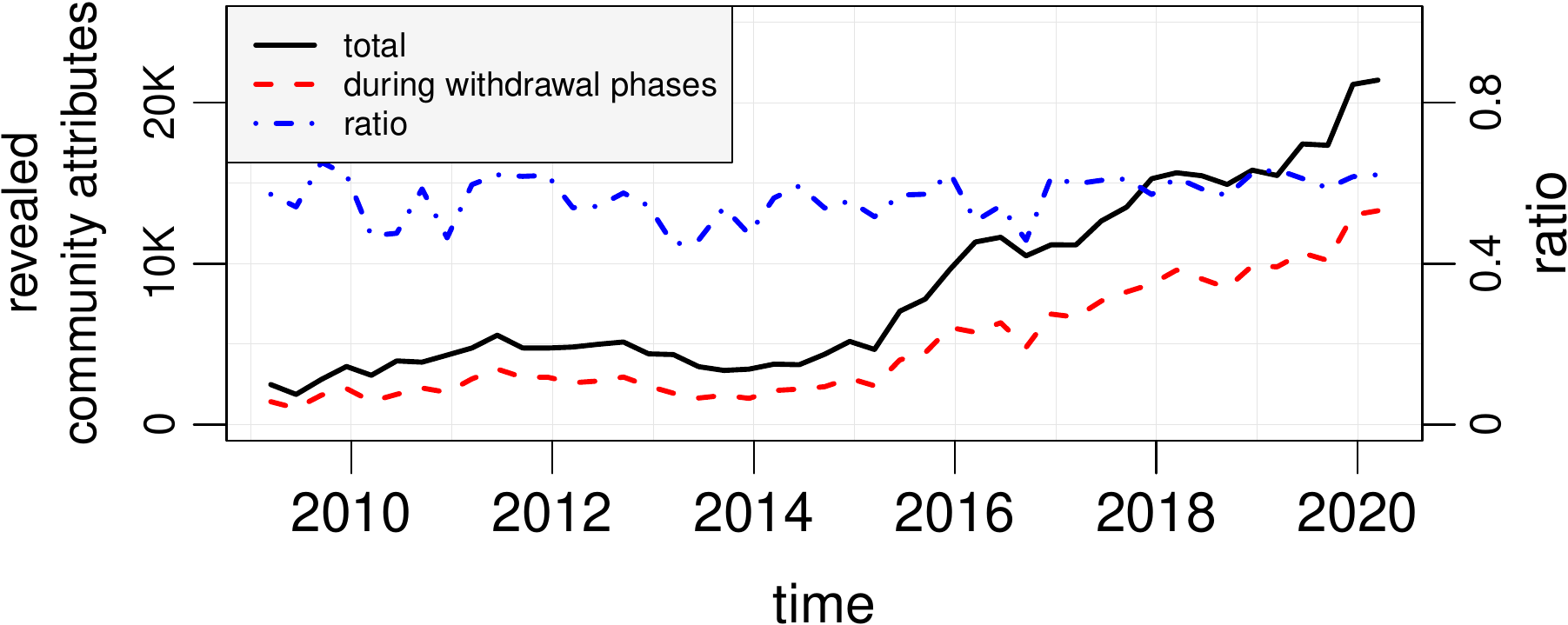}
\caption{Revealed unique community attributes during withdrawal phases of all \ripe beacon prefixes over time.}
\label{fig:communityChanges.longall.phases.overTime}
\end{figure}

\textbf{Revealed Information.}
We have shown that geo-tagging can lead to bursts of announcements just updating
the community attribute, which can lead to re-announcements by neighboring ASes,
in a lab experiment and in the wild.
Given the information hiding character of BGP, we next investigate how
community exploration impacts the amount of information that is
revealed by
community attributes.
We utilize the fixed announcement and
withdrawal phases of the beacon prefixes, and label all announcements
~$\in$~\databeacon according to their appearances in any of the predefined phases,
or outside them. We consider all announcements that appear withing 15 minutes
of the respective phase begins, e.g., between 2:00 to 2:15 UTC for the first
withdrawal phase.

In March 15, 2020, we identify a total of 21,398 unique community attributes.
62\% of all community attributes are revealed exclusively during the
withdrawal phases. Only 17\% are revealed during the announcement phases and
$<$1\% outside both phases. The remaining attributes show up ambiguously.
Historically, this distribution is stable, as can be seen in
Figure~\ref{fig:communityChanges.longall.phases.overTime}. While the number of
unique community attributes per day during withdrawal phases increased multifold
in the last ten years, so did the total number, resulting in a stable ratio
of about 60\%.
A reason for this high number is the global increase in connectivity between
and within ASes and thus more alternative routes are explored during the
withdrawal intervals.
We note that tagged prefix aggregates are not reachable during global
withdrawals.

\begin{figure}[t]
\centering
\includegraphics[width=1.0\linewidth]{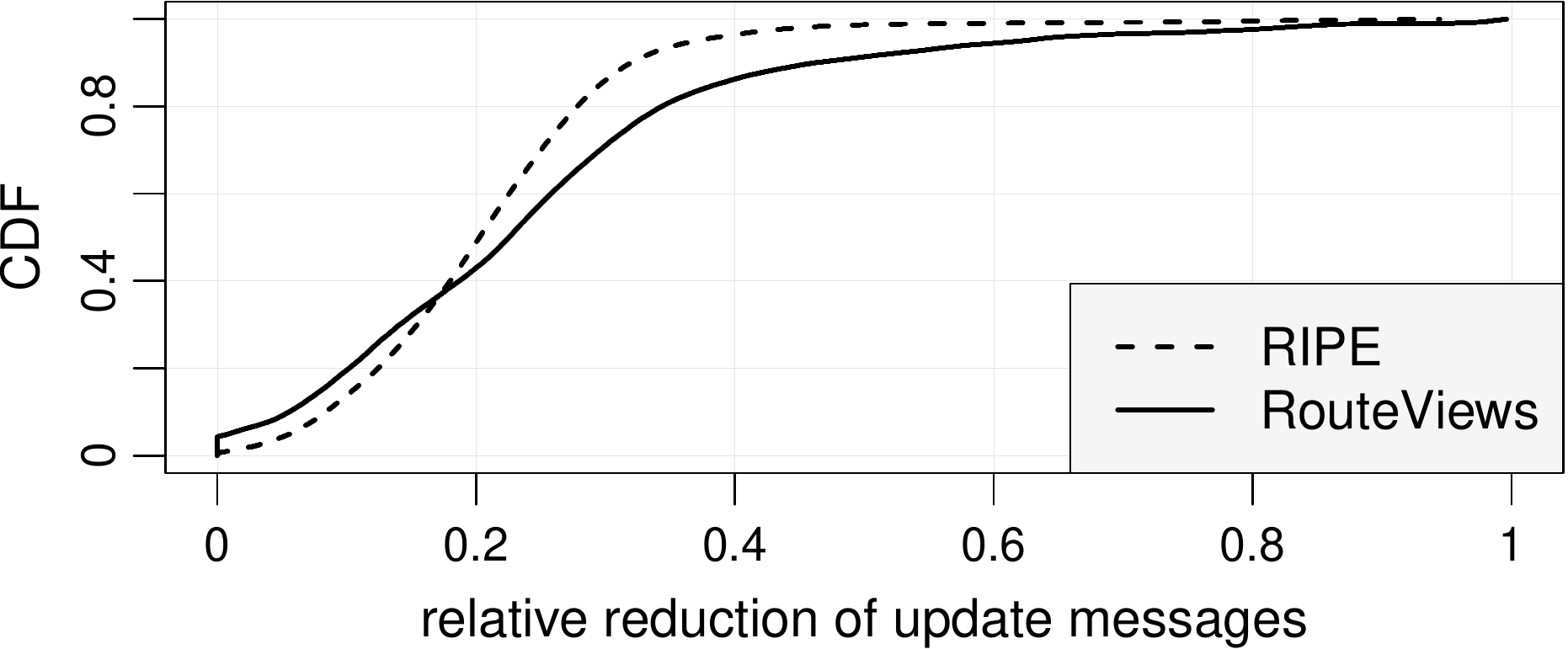}
\caption{Update message reduction of all MRT update files used in this work,
across all time and all collectors.}
\label{fig:communityChanges.mrtReduction.all}
\end{figure}
\textbf{Update message reduction.}
Another aspect to consider about unnecessary announcements is their impact on
message archival.
In order to investigate the update message overhead caused
by unnecessary announcements, we return to our source data, i.e.,
all (251,493) update MRT files from which we have generated the data sets
\datahist, \datanow and \databeacon.
For each individual update file, we parse all the updates
and check if they contain unnecessary announcements.
Note that a single BGP update can contain multiple prefix announcements,
as well as withdrawals for the same AS path.
Given a single update file, we discard any update that contains only
unnecessary announcements, e.g., no path changes or withdrawals.
Since it is difficult to attribute a single prefix to the update message size,
we conservatively count the full update message, even if it includes some
unnecessary announcements.

Figure~\ref{fig:communityChanges.mrtReduction.all} shows the relative reduction
of update messages over all MRT update files.
We note that the distribution does not depend on the time or the collector.
Still, since \rviews and \ripe use a different a binning
(96 and 288 update files per day, respectively)
to archive the updates, here we distinguish between them.
Around 50\% of all \ripe update files are reduced by 20\%+ update messages.
While \rviews show a slightly higher reduction effect (mean=24.8\%),
the overall reduction is comparable to \ripe (mean=20.8\%).
We note that the uncompressed file size correlates positively with the number of
updates message it includes. Their ratio is almost constant for all update files
per day, per collector (with a $\sigma$ of less than 0.05). Thus, we conclude
that by removing updates containing only unnecessary announcements, the
uncompressed data volume of update files can be reduced by 20-25\%.

\begin{figure}[t]
\centering
\includegraphics[width=1.0\linewidth]{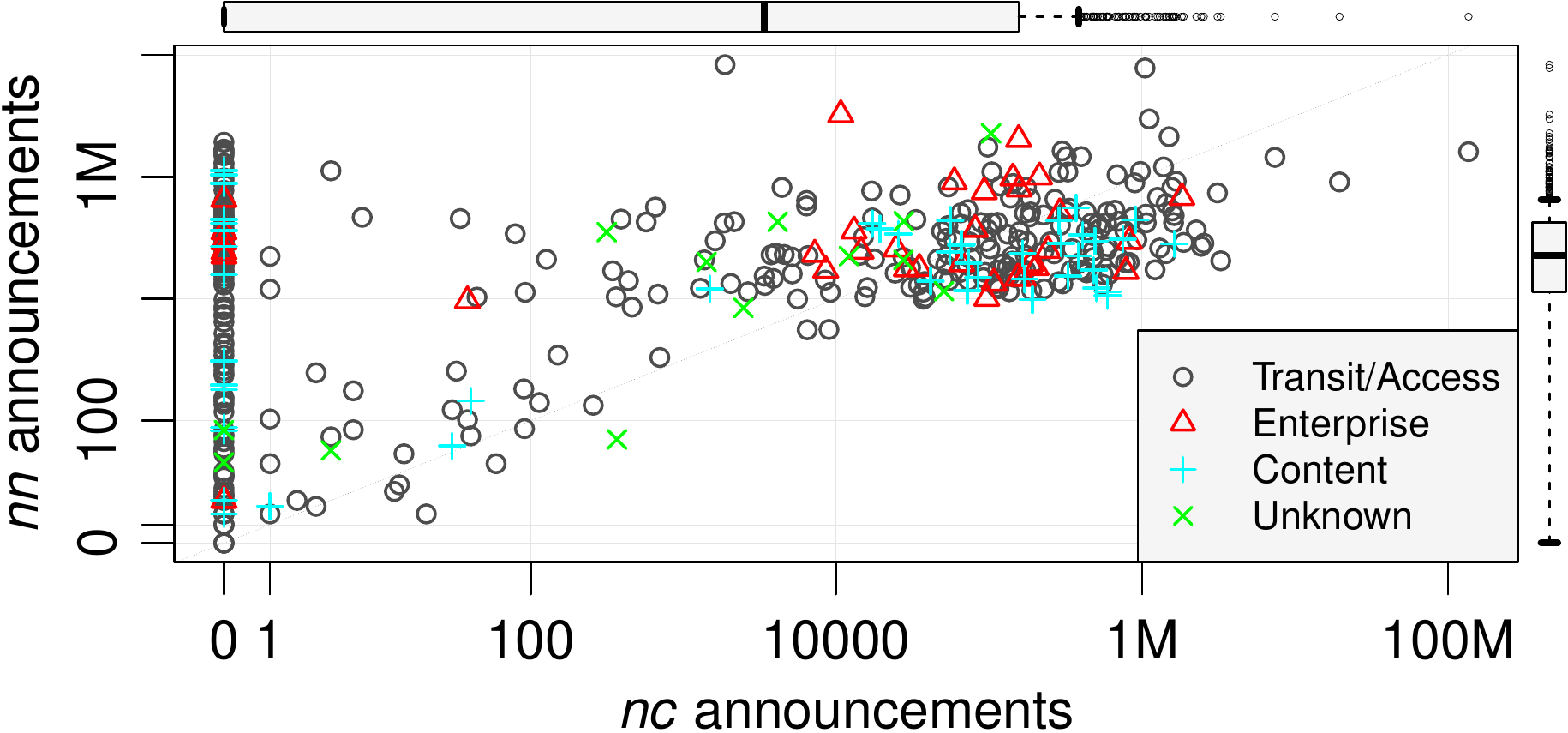}
\vspace{-4mm}
\caption{Unnecessary announcement (\textit{nc} vs. \textit{nn}) behavior of peer ASes in \datanow. No correlation with AS type.}
\label{fig:communityChanges.long-ext.20200315.badASes.as2types}
\vspace{-1em}
\end{figure}
\textbf{Peer AS characterization.}
Next, we take a closer look at ASes that send unnecessary updates, \ie,
\textit{nn} and \textit{nc} announcements.
Thereby, we focus only on ASes peering with collectors, since they provide
the only ground-truth information in our data set.
We note that not all peer ASes send updates on a given day.
From the 581 peers in \datanow, we select 451 that actually send updates
to collectors.
Figure~\ref{fig:communityChanges.long-ext.20200315.badASes.as2types} shows the
number of \textit{nc} announcements (x-axis) and \textit{nn} announcements
(y-axis) for those peers (both axes in log scale).
We observe that while 162 (36\%) peers send \textit{nn} announcements only
(x=0), the majority of peers (285 / 63\%) send both, \textit{nc} and
\textit{nn} announcements. Only 4 peer ASes send no unnecessary announcements
(x and y=0).
Interestingly, the cluster between 10K and 1M indicates a duality in the
behavior of ASes:
there are peers that send 10K+ of \textit{nn} announcements only (x=0),
and there are peers that send 10K+ of both, \textit{nn} and \textit{nc}
announcements.
In fact, there is not a single peer that sends \textit{nc} announcements only,
implying that when communities are forwarded duplicates are always involved.
We believe this is the result of different community propagation and filtering
behavior of the peer ASes.

We further assign an AS type to each of the peers by utilizing CAIDAs as2types
data set \cite{caida:as2type}: We identify 355 Transit/Access, 49 Content, and 34 Enterprise
ASes; for 13 peer ASes no type is available. We see that the different AS types
spread across both dimensions. Thus, we conclude that the AS type does not
correlate with the propagation and filtering behavior of ASes.

%% file: sections/discussion.tex
\section{Discussion} \label{sec:discussion}

As the Internet's core inter-domain routing protocol, the BGP has been extensively studied.  While previous work has
found duplicate updates in the wild~\cite{labovitz1998internet} and identified potential
causes~\cite{park2010investigating,hauweele2018parrots}, we show that BGP communities play a large role in the
generation of unnecessary updates.  First, as a transitive property of BGP messages, communities can induce updates to
propagate through the entire routing system even when the path information is unchanged, the routing decision algorithm
is unaffected, and the receiving AS does not recognize the community.  Second, even when communities are filtered by an
intermediate AS, common implementations still generate a duplicate update, just without the community.  While duplicate
updates without communities do not continue to propagate, we show that they represent a sizable fraction of BGP messages
seen at route collectors and are unnecessary traffic.

Due to the surprising scale of our findings, we were interested to know if
networkers and developers are aware of the behavior we observe and
whether they consider it undesired.
As part of our validation process, we created a website 
documenting our research efforts, including case studies, laboratory
experiments, and open questions directed to the
community~\cite{url:communityexploration}.
We published this website on various mailing lists
(NANOG~\cite{url:nanog.list},
BIRD~\cite{url:bird.list},
FRRouting~\cite{url:frr.list},
and OpenBGPD~\cite{url:bgpd.list})
and got into contact with various developers of routing daemons
and two vendors.
We have also directly contacted two operators of ASes involved in
the propagation of unnecessary announcements, one of which, a large Tier-1 ISP,
responded and confirmed one of our findings that only geolocation communities
are forwarded to, e.g., customer ASes, while other communities are filtered.
While the reaction of the community was affirmative of our findings,
we highlight two perspectives on the generation of unnecessary updates.
First, developers care about default configurations and to what degree they
should predefine the behavior of a router. Indeed, as we show in
laboratory experiments (\S\ref{sec:controlledexperiments}) the default behavior
deviates among different implementations. Also, through internal E-Mail
communication we learn that even among the same vendor different teams of
developers work separately on different implementations.
Second, network operators need to weigh off between memory usage due to state
keeping and the CPU overhead of processing unnecessary update messages. We find
that a system wide increase of processing due to communities has not been
anticipated by the community.

Prior work has shown that the lack of filtering and widespread propagation of BGP communities can leak information about
networks' operation and practices~\cite{SIGCOMM2017-peering-infrastrutcure-outages, IMC2017-blackholing}
and peering~\cite{Inferring-Complex-AS-Relationships:IMC2014,CoNEXT2015-P2F}, and can even be exploited to attack the
routing system~\cite{IMC2018-Communities-attacks}.  Our findings in this work demonstrate an additional motivation for more
rigorous community filtering: reducing unnecessary duplicate update traffic.  Not only does the unnecessary traffic
impact router load and convergence times~\cite{agarwal2004impact}, it increases the load and storage requirements of
systems that monitor BGP traffic including route collectors.
We show that by removing unnecessary announcements from update messages archived
by \ripe and \rviews at least 20\% of data volume can be saved, (\S\ref{sec:analysis}).
As the global use of communities increases and ASes become increasingly
interconnected, the impact of not filtering will place even more strain on the
system.

However, we note several other implications of our findings that we plan to study in future work.  First, communities
are somewhat paradoxical to BGP's emphasis on scalability and information hiding.  For instance, the updates we observe
often allow us to remotely infer the number of interconnections between two ASes and the location where they peer.  Second, from
observing updates and lack of updates at multiple points in the network, we can make rough guesses as to the way
different ASes handle communities. Using more sophisticated network tomography techniques, we plan to classify per-AS
community behavior, for instance those that tag, filter, and ignore.
We hope to use this information to estimate the contribution of ASes that do not
peer with collectors to the generation of unnecessary announcements.
Finally, we believe that communities can enrich
our understanding of anomalous behavior in the routing system beyond existing approaches.  By characterizing the way
ASes observe and process communities, our work provides a first step toward predicting anomalous communities.